\documentclass[onecolumn,showpacs,preprintnumbers,amsmath,amssymb]{revtex4-1}

\usepackage{mathrsfs}
\usepackage{amsfonts}
\usepackage{graphicx}
\usepackage[latin1]{inputenc}
\usepackage{amsmath}
\usepackage{amsfonts}
\usepackage{amssymb}
\usepackage{amsthm}
\usepackage{amsxtra}
\usepackage{fancyhdr}
\usepackage{mathrsfs}
\usepackage{amsmath,amssymb,amsmath}
\usepackage{graphicx}
\usepackage{dcolumn}
\usepackage{bm}
\newcommand{\be}{\begin{equation}}
\newcommand{\ee}{\end{equation}}
\newcommand{\bc}{\begin{center}}
\newcommand{\ec}{\end{center}}
\newcommand{\ben}{\begin{eqnarray}}
\newcommand{\een}{\end{eqnarray}}

\newcommand{\ket}[1]{\left| #1\right>}

\newcommand{\braket}[2]{\left< #1 \right|\left. #2 \right>}

\newcommand{\cor}[1]{\left[ #1 \right]}
\newcommand{\pare}[1]{\left( #1 \right)}
\newcommand{\key}[1]{\left\{ #1 \right\}}

\begin{document}
\title{Statistical measures of complexity for quantum systems with continuous variables}

\author{Daniel Manzano}
\affiliation{Institut f\"ur Theoretische Physik, Universit\"at Innsbruck
Technikerstr. 25 6020 Innsbruck, Austria}
\affiliation{Institut f\"ur Quantenoptik und Quanteninformation
Technikerstr. 21a, 6020 Innsbruck, Austria}
\affiliation{Instituto Carlos I de Fisica Teorica y Computacional, 
University of Granada, Av. Fuentenueva s/n, 18071 Granada, Spain}

\begin{abstract}
The Fisher-Shannon statistical measure of complexity is analyzed for a continuous manifold of quantum observables. 
It is shown that evaluating this measure only  in the configuration or in the momentum spaces does not provide an adequate characterization
  of the complexity of some quantum systems. In  order to obtain a more complete description of complexity two new measures,
  respectively based on the minimization and the integration of
the usual Fisher-Shannon measure over all the parameter space, are proposed and 
compared. Finally, these measures are applied to the concrete case of a free particle in a box.
\end{abstract}

\maketitle
\section{Introduction}

Complexity is a very interesting concept. Even if almost everyone has an intuitive conception of what
complexity is an accepted mathematical definition of it is still not accepted. One reason is that
this is a term that applies to very different kinds of problems, like in computer science 
\cite{kolmogorov:pit65}, ecology and sociology \cite{holling:01} or quantum computing 
\cite{mora:ijqi07}.

In recent years new mathematical measures of complexity for quantum systems with continuous variables 
have been proposed. The principal approach to this problem is to define a measure of complexity for
a probability distribution function (pdf in brief) and then apply it to the pdf of certain variables of the 
system, principally the configuration ($\hat{x}$) or momentum ($\hat{p}$).

One of the most used measures of statistical complexity applied to quantum systems with continuous 
variables is the Fisher-Shannon measure \cite{sen_11,sanudo:jpa08}. The principal condition of this measure 
as a statistical measure of complexity is that it 
has its minimum value for the two extreme probability distributions, the absolutely ordered system (a 
Dirac delta probability distribution) and the absolutely disordered one (a highly flat distribution). These 
quantities have been used in the fields of non-relativistic \cite{dehesa:epjd09,lopezrosa:pa11,angulo:jcp08,angulo:pla08} and
relativistic \cite{manzano:epl10} atomic physics and molecular chemistry \cite{lopezrosa:pccp10}, for example. Other similar
measures are the LMC \cite{lopezruiz:pla95,anteneodo:pla96,catalan:pre02} and the Cramer-Rao \cite{sen:pra07,cover_91},
both having their minimal values for the two extreme distributions. 

One important feature of these measures is their dependence on the space where they are applied. It is known that
they give different results if they are calculated for the configuration or momentum pdfs and 
that sometimes one of these representations gives more information than the other one \cite{angulo:jcp08}, it being not
trivial to realize which one will give the information for a certain state. That fact 
makes it necessary to check both spaces to obtaining a proper analysis of the system. Because of this reason it is complicated 
to say that these measures represent the complexity of the system itself but they represent the complexity
of a determinate observable of the system. It is also clear that very different quantum states can give the same 
Fisher-Shannon measure for a concrete representation. Only in the limit where the pdf characterizes completely the 
state can this measure be considered an intrinsic property of the system itself. 

The purpose of this paper is to analyze the dependence of the Fisher-Shannon measure when we change the basis of the system in a 
continuous way, and the proposal of two generalizations of it that are independent of the basis. The structure of the paper is 
the following: in section \ref{analysis} the dependence of the Fisher-Shannon measure with the selected basis is analyzed, in 
section \ref{sec:gm} we propose and justify two new measures of the complexity of a quantum system that are not calculated for an
 specific space, in section \ref{sec:box} these new measures are applied to the concrete case of a particle in a box, and finally in 
section \ref{sec:conclusion} some conclusions are given.

\section{Analysis of the basis-dependence of the Fisher-Shannon measure}
\label{analysis}

It is well known that position and momentum are not the only two observables that can be defined in a quantum system. 
If a quantum system is defined by the state $\ket{\psi}$ and we have a pair of one dimensional continuous observables 
$(\hat{x},\hat{p})$ (for simplicity we will work with one-dimensional systems, but
all this reasoning can be easily extended to $D$-dimensional systems) 
that are canonical conjugated, the new operators defined by the unitary transformation

\be\label{eq:unitary}
\left(
\begin{array}{c}
\hat{x}_\theta\\
\hat{p}_\theta\\
\end{array}
\right)=\left(
\begin{array}{cc}
\cos \theta & -\sin \theta\\
\sin \theta & \cos \theta\\
\end{array}
\right)
\left(
\begin{array}{c}
\hat{x}\\
\hat{p}\\
\end{array}
\right)
\ee
are also canonical conjugated. It is easy to check that $\hat{p}_\theta$ is equivalent to $\hat{x}_{\theta+\frac{\pi}{2}}$, 
so all the possible operators can be obtained by changing $\theta$ in the interval $\left[0,\pi\right[$. We have a 
continuous manifold of observables defined by the operators

\be
\hat{s}_\theta = \cos \theta\; \hat{x}- \sin \theta\; \hat{p}\qquad \theta \in \left[0,\pi\right[.
\ee 
where $\hat{s}_0$ corresponds with the position and $\hat{s}_{\frac{\pi}{2}}$ with the momentum of the system. 
We use the notation $\ket{s_\theta}$ for the eigenvectors of the operator $\hat{s}_\theta$ with eigenvalue 
$s_\theta$.

For any state $\ket{\psi}$ if we know the wavefunction for a certain parameter $s_\theta$ 
($\braket{s_\theta}{\psi}$)  we can calculate the wavefunction 
for any other variable $s_{\theta'}$ just by the scalar product

\be\label{eq:trans}
\braket{s_{\theta'}}{\psi}=\int_{-\infty}^{\infty} \braket{s_{\theta'}}{s_\theta}\braket{s_\theta}{\psi} ds_{\theta}.
\ee

Finally, the probability distribution for the output of the measurement of $s_\theta$ 
can be determined in the usual way

\be
\rho(s_\theta)=\left| \braket{s_{\theta}}{\psi} \right|^2
\ee

The Fisher-Shannon complexity in the space determined by the observable $\hat{s}_\theta$ is defined by the product
\be\label{eq:fishershannon}
C_{FS}[\rho(s_\theta)]:=I[\rho(s_\theta)] \times J[\rho(s_\theta)],
\ee
where 

\ben
I[\rho(s_\theta)]&=&\int_{-\infty}^{\infty} \rho(s_\theta) \cor{ \frac{d}{dx}\log(\rho(s_\theta)) }^2 ds_\theta \nonumber\\ 
J[\rho(s_\theta)]&=&\frac{1}{2\pi e} \exp\pare{2S[\rho(s_\theta)]},
\een
are the Fisher information and the entropic power of Shannon entropy, respectively. The 
Shannon entropy is defined as $S[\rho(s_\theta)]:=-\int_{-\infty}^{\infty} \rho(s_\theta) \log {\rho(s_\theta)} d s_\theta$. 
 The Fisher-Shannon measure is composed of the product of a measure of the spreading of the function, the
  Shannon entropic power $J[\rho]$, 
 and a measure of possible oscillations of it,
the Fisher information $I[\rho]$ \cite{cover_91,dehesa:epjd09}. All these quantities are based on integrating  
over the parameter $s_\theta$ and their dependence with the parameter $\theta$ is highly non-trivial.

The motivations for extending the analysis of the complexity of quantum systems to this continuous manifold of observables
are two. First, as we will see, for certain systems the densities of $\hat{x}$ and $\hat{p}$ are identical, but 
very different to the densities of $s_{\theta}$ for a general value of $\theta$, that means that measuring this quantity
only in the configuration and momentum spaces will give us a biased information. Second, in some kinds of systems, like in 
quantum optics, the observables defined by Eq. (\ref{eq:unitary}) are measurable in a simple way \cite{paul_04} 
so there is no reason for giving more importance to some of the observables than the others.  The use of a measure 
defined only by the pdf of one of the observables of the system is justified if the purpose is to characterize the complexity 
of this concrete magnitude and not the system itself.

For testing the variation of $C_{FS}[\rho_\theta]$ with the parameter $\theta$ we will use the base defined by the eigenvalues 
of the harmonic oscillator $\key{\ket{n}}_{n=0}^\infty$ ($m=\hbar=\omega=1$) in the configuration space: 

\be
\braket{s_0}{n}=\sqrt{\frac{1}{2^n n!\sqrt{\pi}}} e^{-\frac{s_0^2}{2}} H_n(s_0).
\ee 
In this case the projection (\ref{eq:trans}) to the space defined by $\hat{s}_\theta$ gives

\be
\braket{s_\theta}{n}=\braket{s_0}{n} e^{in\theta},
\ee 
so the density of probability of the elements of the base does not change and the complexity is the same for any $\theta$. 
In Table 1 the values for the first 10 elements of the base are displayed. In a more general framework a quantum state will 
be a linear combination of elements of this base with norm equal to one.  When more than one element are involved that 
situation changes dramatically and there is a highly non-trivial dependence of $C_{FS}$ with $\theta$. As an example let 
us take the following state

\begin{table}
\bc
\begin{tabular}{|c|c|}
\hline
State & $C_{FS}$\\
\hline
$\ket{1}$ & 5.15\\
\hline
$\ket{2}$ & 11.7\\
\hline
$\ket{3}$ & 20.5\\
\hline
$\ket{4}$ & 31.3\\
\hline
$\ket{5}$ & 44.2\\
\hline
$\ket{6}$ & 59.0\\
\hline
$\ket{7}$ & 75.7\\
\hline
$\ket{8}$ & 94.3\\
\hline 
$\ket{9}$ & 114\\
\hline
$\ket{10}$ & 137\\
\hline
\end{tabular}	
\ec
\caption{Values of the Fisher-Shannon complexity for the first ten eigenvalues of the harmonic 
oscillator.}
\end{table}

\be\label{eq:state1}
\ket{\phi_1(a)}=a\ket{0}+\sqrt{1-a^2}\ket{2}\quad a\in \mathbb{R}
\ee
for which the Fisher-Shannon measure gives contradictory results if it is measured only in the configuration ($\theta=0$)  and momentum 
($\theta=\frac{\pi}{2}$) spaces. If $a\equiv a^+=\frac{1}{\sqrt{2}}$ the complexities are $C_{FS}[\rho(s_0)]=2.32$ and 
$C_{FS}[\rho(s_{\frac{\pi}{2}})]=2.95$; on the other hand for $a\equiv a^-=-\frac{1}{\sqrt{2}}$ this measure changes the roles of the 
configuration and momentum spaces. That transition can be easily visualized in Figure 1. By analyzing the behavior of the complexity as a
function of $\theta$ we realize that the states defined by  $a^+$ and $a^-$ are conjugated and they must have the same complexity. 

\begin{figure}[ht!]
\bc
\begin{tabular}{cc}
\includegraphics[scale=0.4]{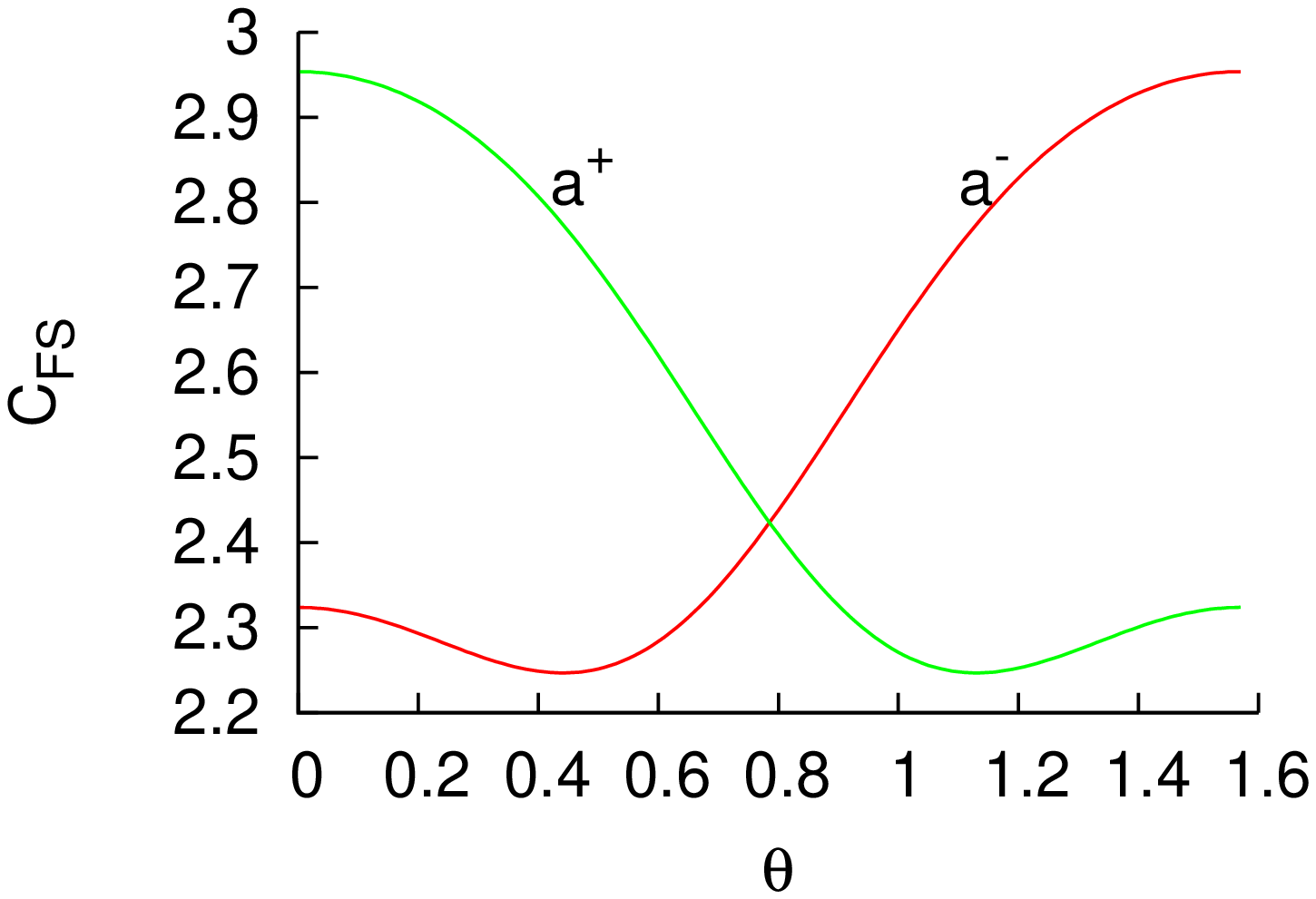} &
\includegraphics[scale=0.4]{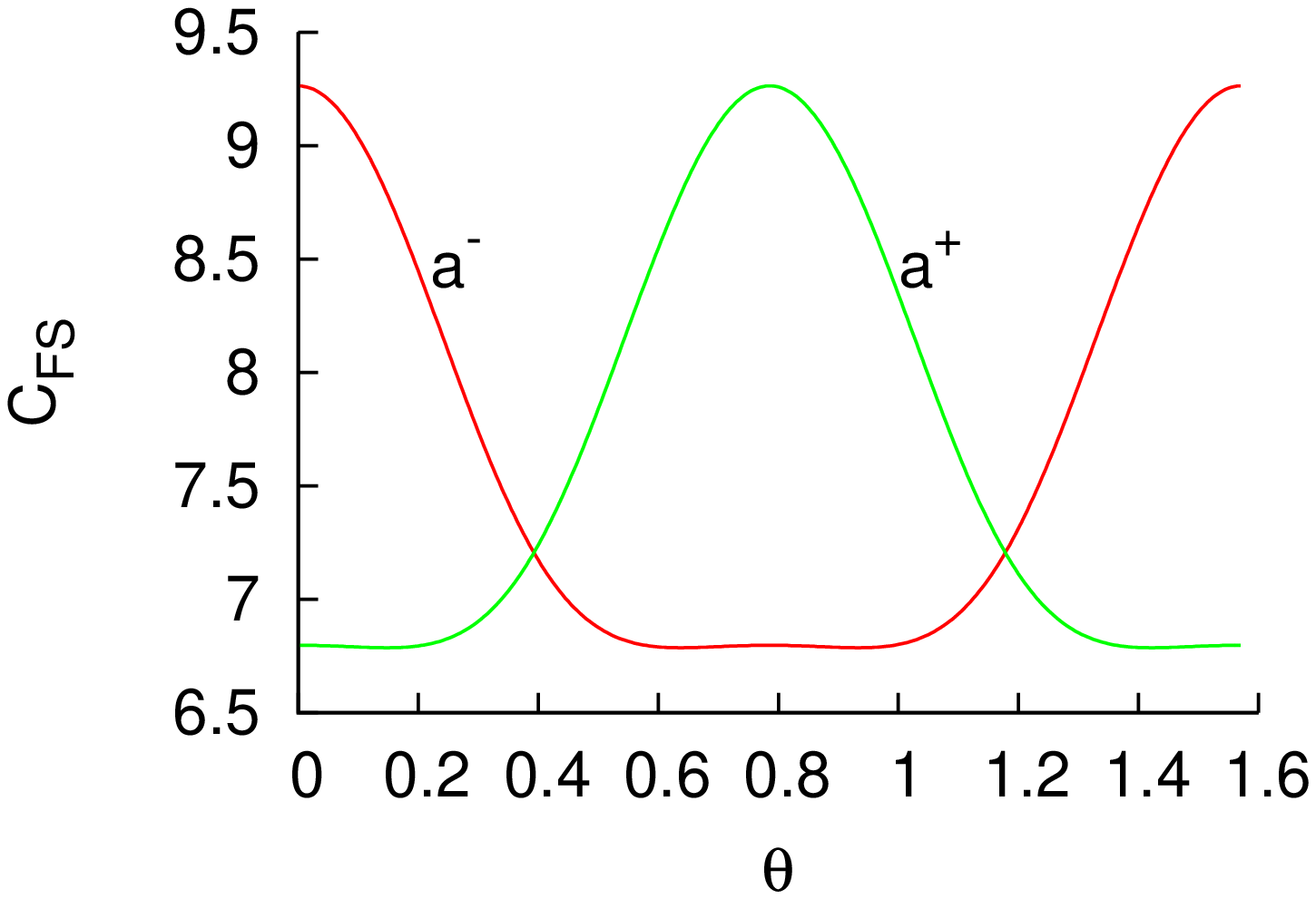}
\end{tabular}
\ec
\caption{(Color online) Representation of the Fisher-Shannon complexity as a function of the parameter $\theta$ for the state 
$\ket{\phi_1}$ (left) and the state $\ket{\phi_2}$ (right) with 
$a^{\pm}=\pm\frac{1}{\sqrt{2}}$.}
\label{fig:fig1}
\end{figure}

The reason for this contradictory results arising from the Fisher-Shannon measure in the configuration and momentum spaces is quite evident. 
The pdfs of position and momentum don't give a complete description of a quantum system \cite{gale:pr68};  only the  full quantum tomogram
containing all such distributions has full information on the state \cite{mancini:pla96,mancini:fp97}.  

Observing the state

\be\label{eq:state2}
\ket{\phi_2(a)}=a\ket{0}+\sqrt{1-a^2}\ket{4}\quad a\in \mathbb{R}.
\ee
For this case with $a^+$ we obtain $C_{FS}[\rho(s_0)]=C_{FS}[\rho(s_{\frac{\pi}{2}})]=6.79763$ and with
$a^-$ they are  $C_{FS}[\rho(s_0)]=C_{FS}[\rho(s_{\frac{\pi}{2}})]=9.26409$, so if we confined our 
analysis to only these observables  we can conclude 
that the state $\ket{\phi_2(a^-)}$ is more complex than $\ket{\phi_2(a^+)}$. That conclusion is not truly real
as can be realized by analyzing the complexity values for all the range of $\theta$. In Figure
2 it is shown that the Fisher-Shannon complexity of $\ket{\phi_2(a^+)}$
has a higher value of the complexity for a continuous interval of the parameter $\theta$, but it is impossible
 to realize it by calculating the complexity only in the configuration and momentum spaces. A more general measure for 
taking into account all the parameter space is required for that.

A usual approach to solve these kinds of problems is to use the  so-called {\it phase space}, that means to work with the
probability density function of $x$ and $p$. In the concrete case of $x$ and $p$ independent that is equivalent to working with the product of 
the complexities in both configurations \cite{angulo:jcp08,angulo_11}. This 
approach would be useful applied to the state $\ket{\phi_1}$, because all the useful information can be obtain in the position 
or momentum spaces. On the other hand, it would give very different values for the system $\ket{\phi_2}$ with  
$a^+$ and $a^-$, not grasping the information about all the intermediate states. 

\section{Global and minimum Fisher-Shannon measures}
\label{sec:gm}

One natural way for generalizing the Fisher-Shannon complexity for taking into account all the parameter space 
is just to integrate over all the possible values of $\theta$. That will be called the global Fisher-Shannon complexity 
($C_{GFS}$), so if we have a state $\ket{\psi}$ and $\rho(s_\theta)=\left|\braket{s_\theta}{\psi}\right|^2$ it 
is defined as 

\be
C_{GFS}[\ket{\psi}]:=\frac{1}{\pi} \int_{0}^{\pi} I[\rho(s_\theta)] \times J[\rho(s_\theta)]  d\theta,
\ee
where the factor $\frac{1}{\pi}$ makes this measure give the same value for the eigenvalues of 
the harmonic oscillator than for the usual measure. 

With this new measure we can analyze again the states of Eqs (\ref{eq:state1}) and (\ref{eq:state2}). Now the results
are independent of the sign of the parameter $a$ 

\ben
C_{GFS}\left[\ket{\phi_1 (a^+)} \right]=C_{GFS}\left[\ket{\phi_1 (a^-)} \right]= 2.53\nonumber\\
C_{GFS}\left[\ket{\phi_2 (a^+)} \right]=C_{GFS}\left[\ket{\phi_2 (a^-)} \right]= 7.63\nonumber\\
\een

Now, as we are using the information about all the continuous pdfs, we can talk of the complexity of the state instead of the complexity of one determinate
observable. As an example in Figure (2) the $C_{GFS}$ for the states $\ket{\phi_1(a)}$ and
 $\ket{\phi_2(a)}$ is shown for $a\in [-1,1]$. 
The $C_{GFS}$ has a quasilinear behavior with $a$ reaching the maximum value when $a=0$, so when the 
state is the eigenvalue of the harmonic oscillator with maximum complexity. 

\begin{figure}[ht!]
\bc
\begin{tabular}{cc}
\includegraphics[scale=0.4]{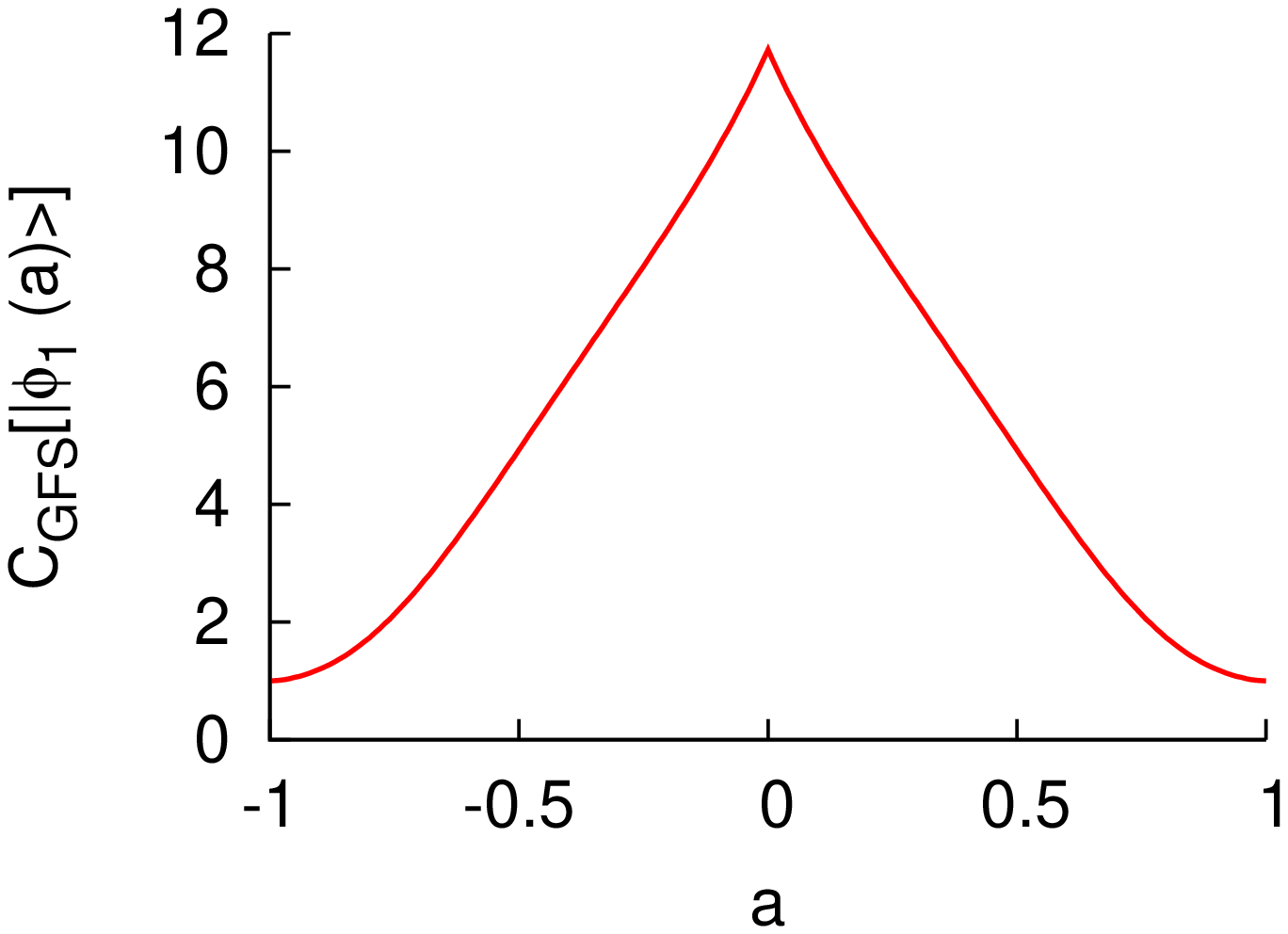} &
\includegraphics[scale=0.4]{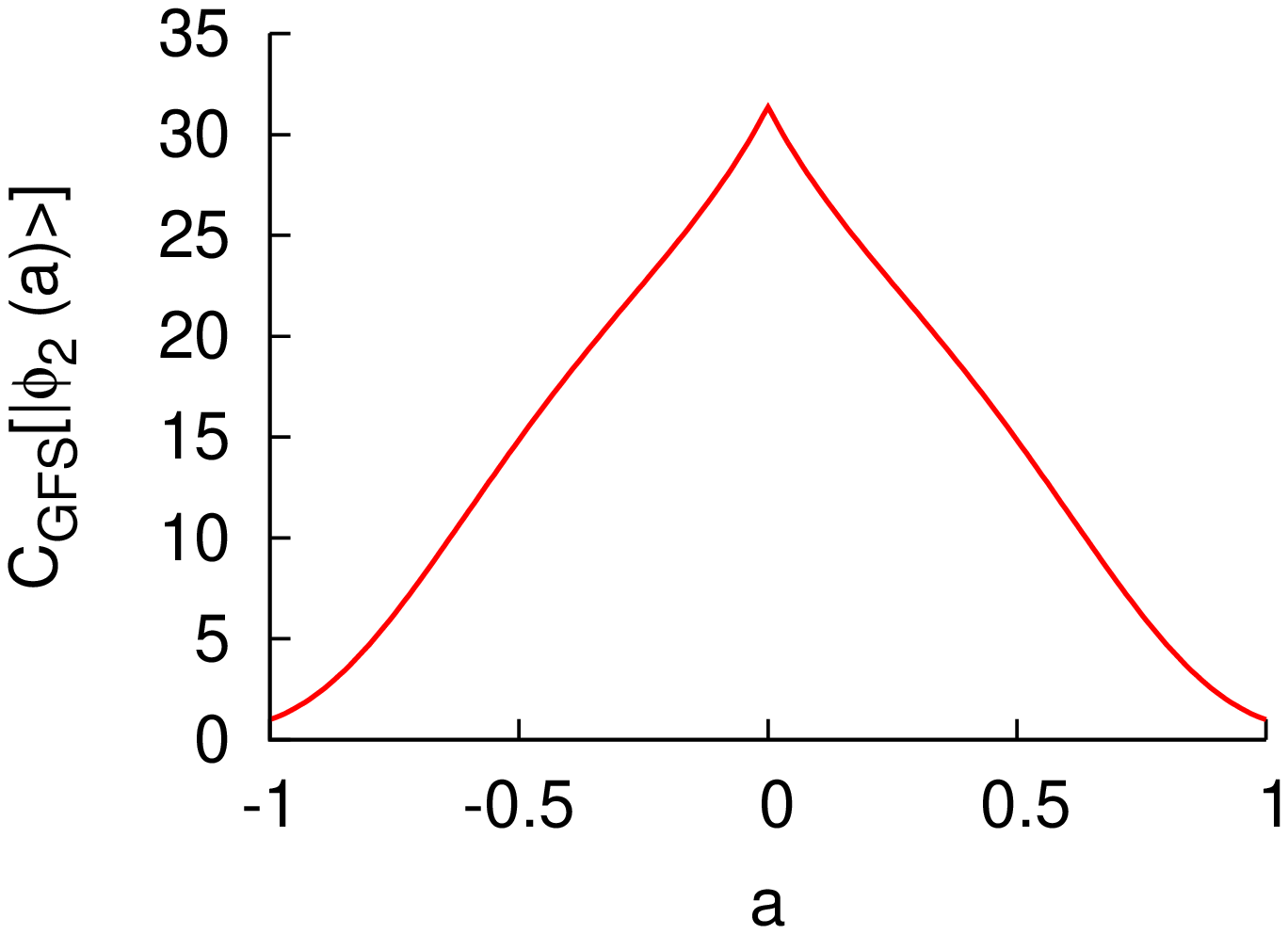}
\end{tabular}
\ec
\caption{(Color online) Global Fisher-Shannon complexity for the state $\ket{\phi_1(a)}$ (left) and $\ket{\phi_2(a)}$ (right) for $a\in [-1,1]$.}
\label{fig:fig2}
\end{figure}

It is clear that with the new measure the usual condition that is imposed for this kind of complexity measure, 
that it must have the minimum value for the Dirac delta and the highly flat distributions, can not be applied. That will
be changed by the following condition.

{\bf Lemma}: The GFS measure of complexity gives the minimum value 
for the states that can be represented as a gaussian distribution for any $s_\theta$. 

{\bf Proof}: If we have a quantum state $\ket{\psi}$ for which probability density is a gaussian distribution in any
 space (for example for the configuration space $s_0$) 

\be
\braket{s_0}{\psi} = \frac{1}{(2\pi\sigma^2)^{\frac{1}{4}}} e^{-\frac{s_0^2}{4\sigma}},
\ee
the application of the projection (\ref{eq:trans}) gives the wavefunction of the state $\ket{\psi}$ for the parameter
$s_\theta$

\be
\braket{s_\theta}{\psi} = \pare{\frac{\sigma^2}{2\pi(\sin^2\theta+\sigma^4 \cos^2\theta)}}
^{\frac{1}{4}} e^{-\frac{s_{\theta}^2 \sigma^2}{4(\sin^2 \theta+\sigma^4 \cos^2 \theta)}}.
\ee

It is clear that if the probability of the measure of $s_0$ in a certain state is a gaussian distribution with 
$\sigma^2$ the distribution of measuring the quantity $s_\theta$ will be gaussian too with a variance 
$\sigma^2_\theta=\frac{\sin^2\theta+\sigma^4\cos^2\theta}{\sigma^2}$. 

By direct integration of the definition (\ref{eq:fishershannon}) for a gaussian distribution
 we obtain that $C_{FS}=1$ independently of the variance. Due to the isoperimetric 
relation \cite{cover_91} $I\times J\ge 1$ (note that our pdfs come from a quantum state and they are normalized to 
unity in all cases), so  it is clear that the Fisher-Shannon complexity has the minimum value for any gaussian 
distribution independently of what observable $s_\theta$ is represented and so the global Fisher-Shannon will be minimal for any gaussian state, 
because it is the integration of a function that always reaches its minimum value. This proof can be trivially extended to the $D$-dimensional problem. 

\vspace{.5cm}
The second possibility for generating a base-independent measure of complexity is to take the minimum value of the usual Fisher-Shannon in the complete 
range of $\theta$. That measure follows the philosophy of Kolmogorov's complexity, that defends that the complexity of a system must be 
 calculated in its simplest description. That will be called the minimum Fisher-Shannon measure of complexity.

\be
C_{MFS}[\ket{\psi}]:=\min_{\theta\in [0,\pi]} I[\rho(s_\theta)] \times J[\rho(s_\theta)].
\ee
This measure must be interpreted as  the minimum complexity of the system for a single variable. It can happen that 
different states have a similar $C_{MFS}$. The interpretation of this fact is that, even if the states are different, 
there is some representation where they share the same complexity. 

We can calculate now the minimum complexity of the states of Eqs (\ref{eq:state1}) and (\ref{eq:state2}). The results
 are similar to the global complexity.

\ben
C_{MFS}\left[\ket{\phi_1 (a^+)} \right]=C_{MFS}\left[\ket{\phi_1 (a^-)} \right]= 2.25\nonumber\\
C_{MFS}\left[\ket{\phi_2 (a^+)} \right]=C_{MFS}\left[\ket{\phi_2 (a^-)} \right]= 6.79 \nonumber\\
\een

\begin{figure}[ht!]
\bc
\begin{tabular}{cc}
\includegraphics[scale=0.4]{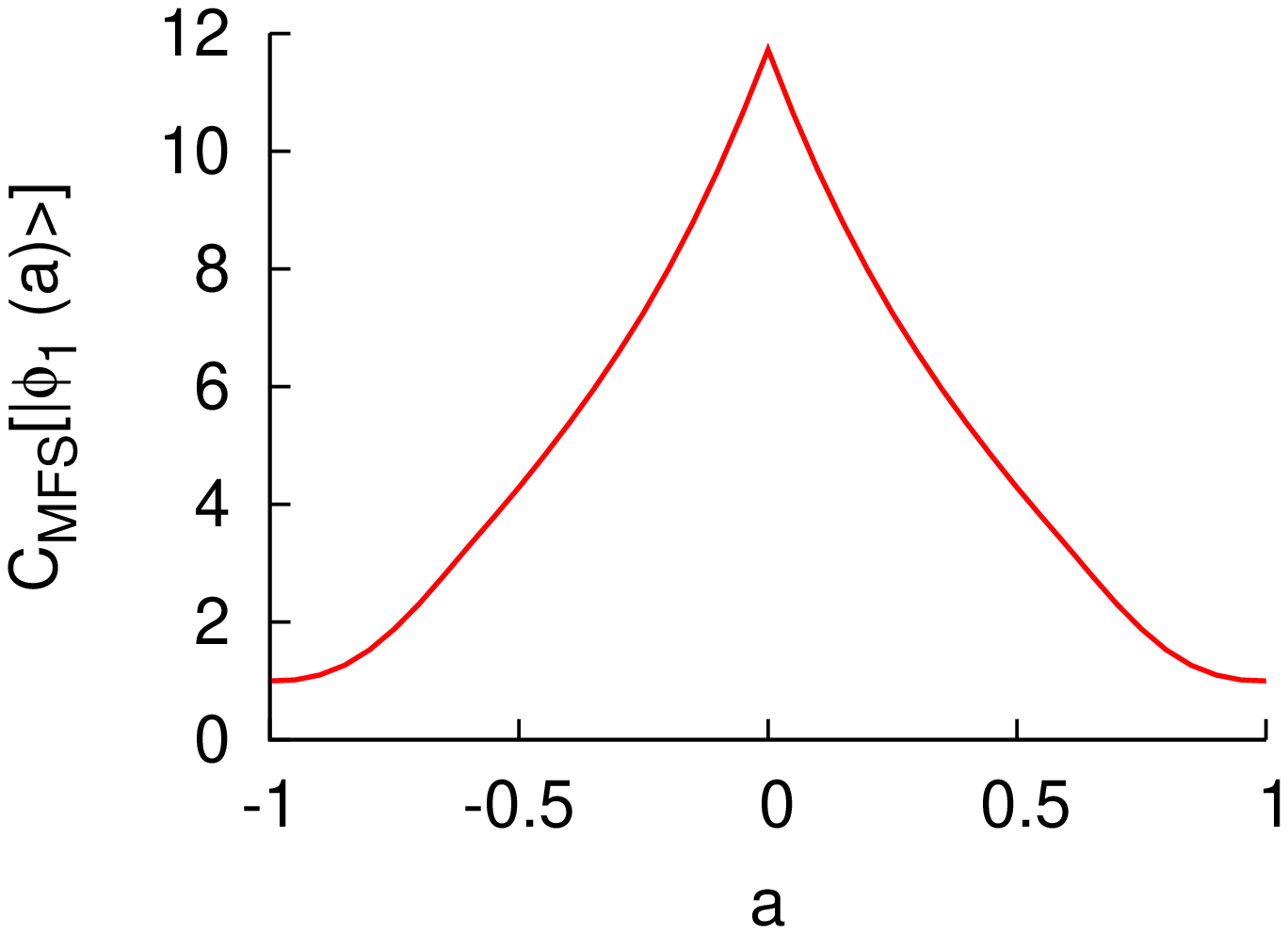} &
\includegraphics[scale=0.4]{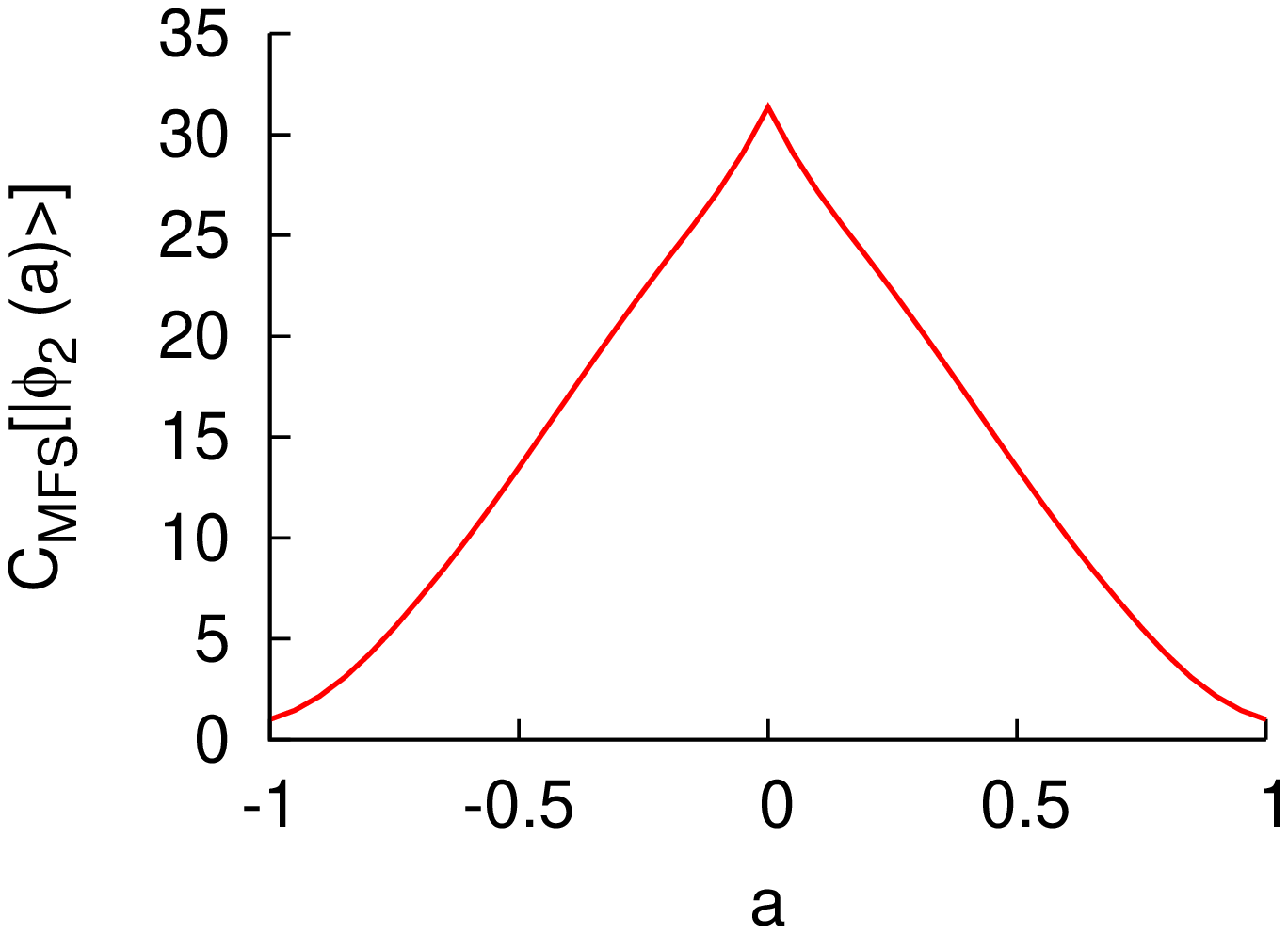}
\end{tabular}
\ec
\caption{(Color online) Minimum Fisher-Shannon complexity for the state $\ket{\phi_1(a)}$ (left) and $\ket{\phi_2(a)}$ (right) for $a\in [-1,1]$.}
\label{fig:fig3}
\end{figure}

With this definition it is possible to analyze the complexity of both states as a function of the parameter $a$ as has been 
plotted in Fig \ref{fig:fig2} for the GFS. The results are really very similar for both measures as can be checked in Fig. 
\ref{fig:fig3}.

Finally for the MFS measure we have a similar condition as we have for the GFS. 

{\bf Lemma}: The MFS measure of complexity gives the minimum value 
for the states that can be represented as a gaussian distribution for any $s_\theta$. 

The proof is equivalent to the global case.

This new definition of complexity, as non-gaussianity, is compatible with the original Fisher-Shannon and LMC measures. These 
measures are usually defined by the constraint that they must be minimal for the two simplest distributions, the Dirac delta and highly 
flat distributions. Following the previous reasoning it is easy to check that these measures also give minimal results for any gaussian distribution, 
so they can be consider as measures of non-gaussianity.

These two measures are intrinsically different even if they give similar results. The global Fisher-Shannon measure 
gives an average measure of the complexity of describing the system in any base. That is useful if we want to take
all the possibilities into account or if we have a system that cannot be easily described in an arbitrary base. 
The minimum Fisher-Shannon is based in Kolmogorov's idea of complexity that states that a system is as complex as 
its simplest description. In both cases we are measuring how far is our system from a gaussian description. In the concrete 
case of being interested in a particular observable one must use the original measures, because a concrete behavior can 
be masked when it is mixed with all the other observables.

\section{Free particle in a box}
\label{sec:box}

As a physical example let us see the concrete case of a one-dimensional particle in a box. This is a quantum system 
with a potential $V(x)=0$ if $x<a$ and an infinite potential otherwise. For simplicity we will work in a concrete 
system with $a=1$. The Fisher-Shannon complexity and other information-theoretical measures of this system have 
been recently analyzed \cite{lopezrosa:jmc11} both in position and momentum spaces. 

The wavefunctions of the energy eigenstates of this system for $\theta=0$ read 

\be
\psi_n(x) = \braket{s_0}{\psi}_n=
\left\{
\begin{array}{cc}
\sin \pare{\frac{\pi n}{2}(x-1)}& |x|\le 1\\
0 & |x|> 1\\
\end{array}
\right.
\ee
and they can be calculated for any value of $\theta$ just by the use of relation (\ref{eq:trans}). For $\theta=0$ the 
Fisher-Shannon measure reads \cite{lopezrosa:jmc11}

\be
C_{FS}[\rho_n(s_0)]=\frac{8\pi n^2}{e^3}
\ee
and for $\theta=\frac{\pi}{2}$ it is

\be
C_{FS}[\rho_n(s_\frac{\pi}{2})]=\frac{\exp (2K(n))}{24 \pi e} \pare{ 1- \frac{6}{\pi^2n^2}},
\ee
where $K(n)$ is the trigonometric integral

\ben
K(n)&=&\log \pare{\frac{8}{\pi}}\\
&&-\pi\int_\frac{\pi n}{2} ^\infty n^2 \frac{\sin^2 (t)}{\pare{t^2+\pi n t}^2} \log \cor{n^2\frac{\sin^2 t}{\pare{t^2+\pi n t}^2}} dt.\nonumber
\een

\begin{figure}[ht!]
\bc
\includegraphics[scale=0.5]{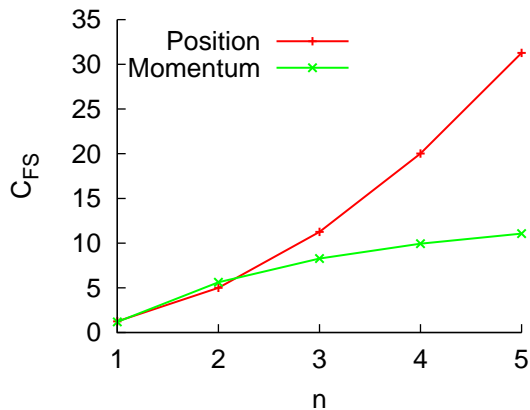} 
\ec
\caption{(Color online) Fisher-Shannon measure of complexity for a particle in a one-dimensional box in position and momentum spaces.}
\label{fig:fig4}
\end{figure}

In Figure \ref{fig:fig4} the complexities of the first five states of this system are plotted. It is clear that these measures of complexity
have a very different behavior when they are applied to one or the other base. 

\begin{figure}[ht!]
\bc
\begin{tabular}{cc}
\includegraphics[scale=0.4]{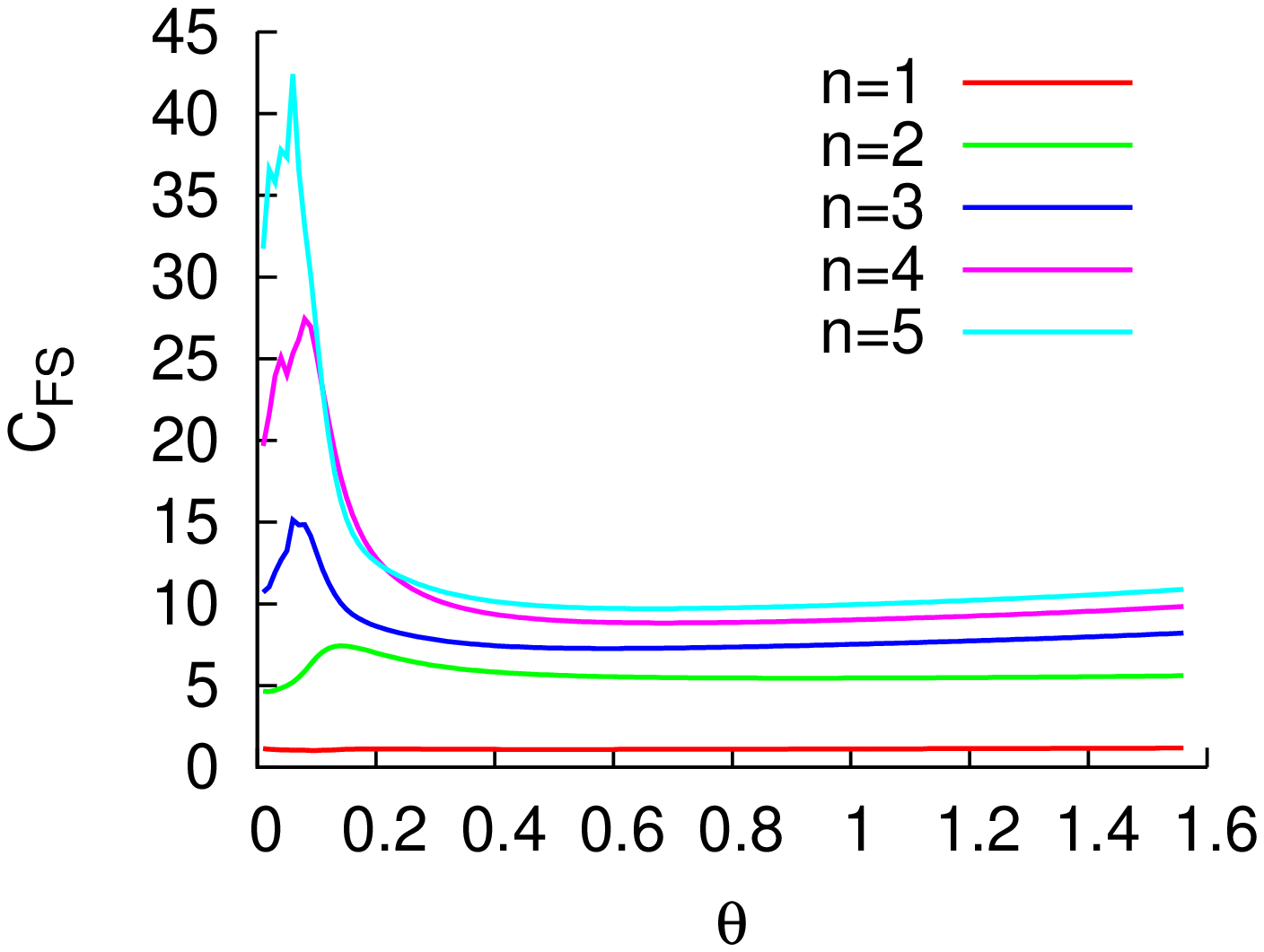} &
\includegraphics[scale=0.4]{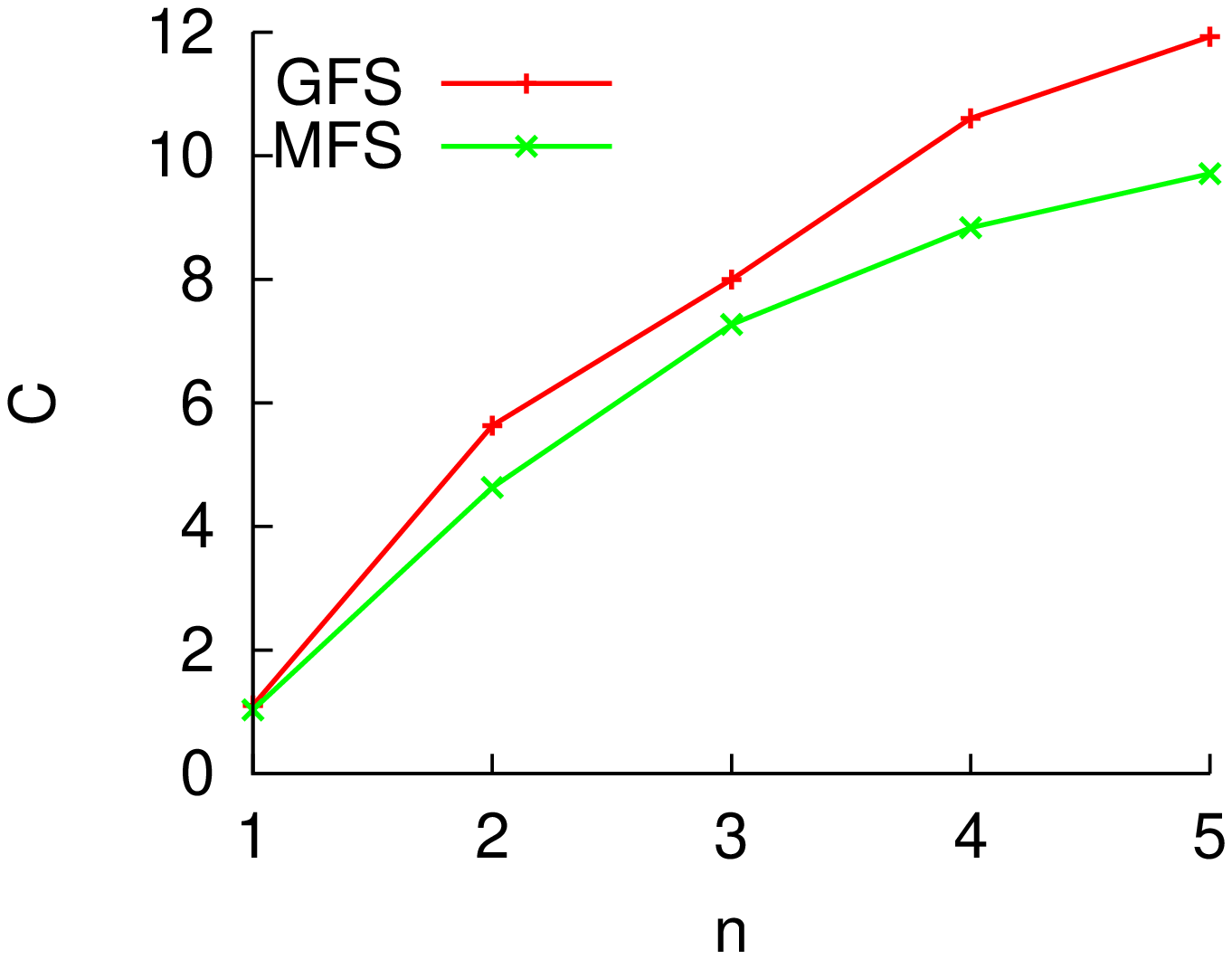}
\end{tabular}
\ec
\caption{(Color online) Left: Fisher-Shannon measure for the first 5 energy eigenstates of a quantum particle in a box as a function of the 
parameter $\theta$. Right: global and minimum Fisher Shannon measures as a function of the quantum number $n$.}
\label{fig:fig5}
\end{figure}

For understanding this behavior we must analyze the Fisher-Shannon measure for all the range of the parameter $\theta$. That measure is plotted 
in figure \ref{fig:fig5} (left) for the first 5 states of this system. It is clear that the behavior of the Fisher-Shannon measure as a function 
of the parameter $\theta$ is  non-trivial, principally for small values of $\theta$. If  $\theta$ approaches to $\pi/2$ the measure becomes
flat and it saturates. We can analyze also the global and minimum fisher-Shannon measure for this system as a function of $n$. That is also 
plotted in Figure \ref{fig:fig5} (right). For this case both measures have a similar and increasing behavior. That is because the global 
measure is principally conditioned by the long and monotone tail that appears for $\theta>0.2$. By comparing both measures we can 
conclude, in general, if the complexity of the system has or not an strong variation for different descriptions. For this concrete system as both 
measures are very similar we can realize that the system is equally complex for a large range of $\theta$.

This example illustrate the utility of this measures. The complexity of the system close to $\theta=0$ is quite different from the rest of 
the parameter space. The usual Fisher-Shannon measure, the GFS and the MFS give complementary and compatible information 
about the system.

\section{Conclusions}
\label{sec:conclusion}
In conclusion, we have analyzed the Fisher-Shannon measure of complexity for a continuous manifold of observables and we realize
that measuring it only in the configuration and momentum spaces does not give complete information. It is clear that for certain 
quantum states this measure can even give contradictory results if it is not measured for all the observables. To avoid this fact
we propose a global measure by the integration over all the parameter space and a second one by the minimization over it.
These measures require changing the condition for statistical measures of complexity by a more general condition. 
These new measures will be more relevant for quantum optic systems, 
where all the observables involved can be measured in an experimental way. In any case, for general quantum systems, both the
the global and minimum Fisher-Shannon measures must be considered more legitimate measures of the system than the usual measures because they do not 
depend on which spaces are calculated. Even if the definitions are made for a one-dimensional system it is trivial to extend it to 
the general case of a quantum system with $D$ continuous variables. 

The physical example of a free particle in a box illustrates how the description of a quantum system can dramatically change if 
it is made for a single parameter or in general. For this concrete case the GFS and the MFS give similar results, but they are very 
different from the Fisher-Shannon measure in position space. 

Finally, let us  point out that this extension has  also been applied to the LMC shape complexity \cite{lopezruiz:pla95}, and to 
the Cramer-Rao measure \cite{sen:pra07} giving the global LMC (GLMC), global Cramer-Rao (GCR), minimum LMC (MLMC) and minimum Cramer-Rao (MCR)
 measures of complexity for a quantum system. These measures have a similar behavior than the GFS and MFS. 
This kind of extension is also possible for other theoretically-based measures like the Jensen-Shannon divergence \cite{antolin:jcp09}. 

The research was funded by the Austrian Science Fund (FWF) and the Junta de Andaluc\'ia, project FQM-165, together 
with the Campus de Excelencia Internacional. The author would like to acknowledge A. Lasanta and S. L\'opez-Rosa 
for useful discussions.


\end{document}